\documentclass[aps,prb,twocolumn]{revtex4}
\usepackage{graphicx}
\usepackage{dcolumn}
\usepackage{amsmath}
\usepackage{amssymb}

\def\uu{\uparrow\uparrow}
\def\ud{\uparrow\downarrow}

\def\dd{\downarrow\downarrow}
\def\rv{{\bf r}}

\def\uv{{\bf u}}

\def\kfr{\frac{k_Fu}{\phi}}
\def\gcav{\overline{g}_c}
\def\gxcav{\overline{g}_{xc}}
\def\ec{\epsilon_c}

\def\be{\begin{equation}}
\def\ee{\end{equation}}
\def\bea{\begin{eqnarray}}
\def\eea{\end{eqnarray}}
\begin{document}
\title{Pair distribution function of the spin-polarized electron gas: \\ 
A first-principles analytic model for all uniform densities}
\author{Paola Gori-Giorgi$^1$ and John P. Perdew$^2$}
\affiliation{$^1$INFM Center for
  Statistical Mechanics and Complexity, and
Dipartimento di Fisica, Universit\`a di Roma ``La Sapienza'', 
Piazzale A. Moro 2, 00185 Rome, Italy \\ 
$^2$Department of Physics and Quantum Theory Group, Tulane
University, New Orleans, Louisiana 70118 USA}
\date{\today}
\begin{abstract}
We construct analytic formulas that represent the
coupling-constant-averaged pair distribution function $\gxcav(r_s,
\zeta, k_Fu)$ of a three-dimensional non-relativistic ground-state electron
gas constrained to a uniform density with density parameter $r_s =
(9\pi/4)^{1/3}/k_F$ and relative spin polarization $\zeta$ over the
whole range $0<r_s<\infty$ and $-1<\zeta<1$, with energetically-unimportant
long range ($u\to \infty$) oscillations averaged out.  The pair distribution
function $g_{xc}$ at the physical coupling constant is then given by
differentiation with respect to $r_s$.  Our formulas are constructed using
{\em only} known theoretical constraints plus the correlation energy
$\ec(r_s,\zeta)$, and accurately reproduce the $g_{xc}$ of the Quantum
Monte Carlo method and of the fluctuation-dissipation theorem with the
Richardson-Ashcroft dynamical local-field factor.  Our $g_{xc}$ is
correct even in the high-density ($r_s\to 0$) and low-density 
($r_s \to \infty$) limits.  
When the spin resolution of $\ec$ into $\uu$, $\dd$, and
$\ud$ contributions is known, as it is in the high- and low-density
limits, our formulas also yield the spin resolution of $g_{xc}$.
Because of these features, our
formulas may be useful for the construction of density functionals for
non-uniform systems.
We also analyze the kinetic energy of correlation into contributions from
density fluctuations of various wavevectors.
The exchange and long-range correlation parts of our 
$\gxcav(r_s,\zeta,k_Fu)-1$
are analytically Fourier-transformable, so that the static structure
factor $\overline{S}_{xc}(r_s,\zeta,k/k_F)$ is easily evaluated.
\end{abstract}
\maketitle
\section{Introduction, definitions, and outline}
The exchange-correlation
pair-distribution function $g_{xc}(\rv,\rv')$
of an $N$-electron system is defined as
\be
g_{xc}(\rv,\rv')=\frac{N(N-1)}{n(\rv)n(\rv')}\int|\Psi(\rv,\rv',\rv_3...
\rv_N)|^2 d\rv_3...d\rv_N,
\label{eq_defg}
\end{equation}
where $n(\rv)$ is the electron density and $\Psi$ is the many-body 
wavefunction. 
Its coupling-constant average $\overline{g}_{xc}(\rv,\rv')$ 
is equal (in the Hartree units used throughout) to
\be
\overline{g}_{xc}(\rv,\rv')=\int_0^1d\lambda\, g_{xc}^{\lambda}(\rv,\rv'),
\end{equation}
where $g_{xc}^{\lambda}(\rv,\rv')$ is the pair-distribution function when the
electron-electron interaction is $\lambda/|\rv-\rv'|$ and the density is
held fixed at the physical or $\lambda=1$ density. The coupling-constant 
averaged $\overline{g}_{xc}$ plays a crucial
role in density functional theory, since it can account for the kinetic
energy of correlation.\cite{Gunreview}
In fact, $n(\rv')\,[\gxcav(\rv,\rv')-1]$ is the density at $\rv'$ of the 
exchange-correlation hole around an electron at $\rv$.

In the uniform electron gas,
$n(\rv)=n$ and $g_{xc}(\rv,\rv')$ only depends
on $u=|\rv-\rv'|$, and parametrically on the density parameter
$r_s=(3/4\pi n)^{1/3}$ and on the spin-polarization 
$\zeta=(N_{\uparrow}-N_{\downarrow})/N$. The coupling-constant average
is in this case\cite{PW} equivalent to an average over $r_s$:
\be
\overline{g}_{xc}(r_s,\zeta,k_Fu)=\frac{1}{r_s}\int_0^{r_s} 
g_{xc}(r_s',\zeta,k_Fu)\,dr_s',
\label{eq_gavdef}
\end{equation}
where $k_F=(9\pi/4)^{1/3}/r_s$ is the Fermi wavevector. Clearly then
\be
g_{xc}(r_s,\zeta,k_F u)=\frac{\partial}{\partial r_s}\left[
r_s\,\gxcav(r_s,\zeta,y)\right]\Big|_{y=k_Fu},
\end{equation}
and
\begin{equation}
g_{xc}^{\lambda}(r_s,\zeta,k_Fu) = g_{xc}(\lambda r_s,\zeta,k_Fu).
\end{equation}
The high-density ($r_s\to 0$) limit is the
weak-interaction limit in which the kinetic energy dominates.
Relativistic effects are important for $r_s \lesssim 0.01$.  The low-density
($r_s \to \infty$) limit is the strong-interaction limit in which
the Coulomb potential energy dominates.  For $r_s \gtrsim 100$, the true 
ground-state density is not uniform,\cite{cepald} but there is still a 
wavefunction that achieves
the lowest energy of all those constrained to a given uniform density.\par

The electron gas of uniform density is a paradigm of the density
functional theory\cite{Gunreview} for real, non-uniform electronic systems.  
The exchange-correlation energy of the uniform gas is the input to the local
spin density approximation, while the coupling-constant-averaged
pair-distribution function is an input to the derivation of
gradient-corrected functionals,\cite{GGA,nxc} to the construction
of the corresponding system-averaged exchange-correlation hole of a
non-uniform density,\cite{nxc} and to the implementation of the fully-nonlocal
weighted density approximation.\cite{WDA,Chacon,RTC}  We hope that our improved
analytic model will be useful for these purposes, and also for the
construction of new and more accurate functionals.  In particular, the
spin-resolved version of our model, when fully developed, could bring
useful new information for the construction of functionals. 
Indeed, simple hypotheses for the spin resolution have
already been used to construct several correlation functionals.\cite{Bec,CH}
The uniform-gas $g_{xc}$ is also relevant to density matrix functional 
theory.\cite{Ziescherodi}\par

The static structure factor $S_{xc}(r_s,\zeta,k/k_F)$ is the Fourier 
transform
\bea
S_{xc}(r_s,\zeta,k/k_F) & = & 1+\frac{4}{3\pi}\int_0^{\infty} 
[g_{xc}(r_s,\zeta,k_F u)-1]\times \nonumber \\
& &  (k_F u)^2 \frac{\sin ku}{ku}\, d (k_F u),
\label{eq_Sxcdef}
\eea
and its coupling-constant average $\overline{S}_{xc}$ is obtained by 
changing $g_{xc}$ into $\gxcav$ in Eq.~(\ref{eq_Sxcdef}). 
Usually $g_{xc}$ and consequently 
$\overline{g}_{xc}$, $S_{xc}$, and $\overline{S}_{xc}$ are divided into 
exchange and correlation contributions:
\be
g_{xc}(r_s,\zeta,k_Fu)=g_x(\zeta,k_Fu)+g_c(r_s,\zeta,k_Fu),
\end{equation}
where the exchange function $g_x$ is obtained by putting a Slater determinant
of Kohn-Sham orbitals (or of Hartree-Fock orbitals) into Eq.~(\ref{eq_defg}).
For a uniform electron gas, both Kohn-Sham and Hartree-Fock orbitals are 
plane waves, and $g_x$ is a simple function of $k_Fu$. 
The exchange-only pair-distribution function does not depend explicitly on 
$r_s$, so that $\overline{g}_x=g_x$: the explicit dependence on $r_s$ 
only appears when Coulomb repulsion is taken into account in the wavefuction.
\par
Both $g_x$ and $g_c$ have long-range oscillations.  At high densities,
these are Friedel oscillations; at low densities, they represent the
incipience of Wigner-crystal order within the liquid phase of uniform
density.  These oscillations are energetically unimportant in the
following sense:\cite{PW} A model which omits them but is constrained to have
the same energy integral can correctly describe the short-range
correlation while averaging out the oscillations of the long-range
correlation.  The energetic unimportance of the oscillations is probably
a consequence of the long-range and ``softness'' of the Coulomb interaction.
\par
Available analytic models\cite{PW,GSB}
of $g_{c}$ and $\overline{g}_{c}$
for the uniform electron gas break down at 
high\cite{Rassolov,comment} ($r_s\lesssim 0.1$) and 
low ($r_s > 10$) densities.
In this paper, we present a new model 
for the nonoscillatory part of $\gcav$ (and hence $g_c$) which fulfills
most of the known exact properties and is valid 
over the whole ($0< r_s< \infty$) density range and for all
spin polarizations $\zeta$. Our model is built up
by interpolating between the short-range part recently computed in
Ref.~\onlinecite{GP} and the long-range nonoscillatory part which is 
exactly given by the random-phase approximation\cite{WP} (RPA).
Exact small-$u$ and large-$u$ expansions are recovered up to
higher orders with respect to currently available models.\cite{PW,GSB}
All the parameters which appear in our interpolation scheme are fixed
by exact conditions. 
We also build up a new nonoscillatory exchange $g_x$ which fulfills exact
short-range and long-range properties up to the same order as
our $\gcav$ does. 
\par
The paper is organized as follows. In Sec.~\ref{sec_prop},
we list the known exact properties of $g_{xc}$ and $\overline{g}_{xc}$,
and the major limitations of the models of Refs.~\onlinecite{PW}
and~\onlinecite{GSB}. We then
present our nonoscillatory model for exchange (Sec.~\ref{sec_ex})
and for correlation (Sec.~\ref{sec_corrhole}). In Sec.~\ref{sec_res},
we discuss our results for exchange and correlation over the whole density
range. At metallic densities, we compare our analytic model with the
available Quantum Monte Carlo (QMC) data,\cite{cepald,OHB} finding
fair agreement (Fig.~\ref{fig_gcQMC}). We also computed $g_c$ 
corresponding to the dynamic 
local-field factors of Richardson and Ashcroft\cite{RA} (RA), in order
to see better how our model averages out the long-range oscillations
(currently not available from QMC). 
In this way, we are also able to show
the effect of a dynamic local-field factor on the long-range oscillations, by
comparing the RA result with the RPA (corresponding to zero local-field
factor) long-range $g_c$ (Fig.~\ref{fig_gLR}). At high density, we find that 
our model is in very good agreement with exact 
calculations\cite{Rassolov,reply}
(Fig.~\ref{fig_gcHD}), and at low density it does not break down
and shows the expected $\zeta$ dependence (Fig.~\ref{fig_LD}).
We also compare (Fig.~\ref{fig_gold}) our model with previous 
models,\cite{PW,GSB} and discuss the qualitative effects of 
correlation (Fig.~\ref{fig_Kxc}).
In Sec.~\ref{sec_spinres}, we discuss how to extend our scheme to the 
spin-resolved ($\uu$, $\dd$ and $\ud$) pair-distribution functions.
The wavevector analysis of the kinetic
energy of correlation corresponding to our $S_c$ and $\overline{S}_c$ is
presented in Sec.~\ref{sec_tc}. Section~\ref{sec_conc} is devoted
to conclusions and perspectives. 

\section{Exact properties, AND LIMITATIONS OF PREVIOUS MODELS}
\label{sec_prop}
We list below most of the known exact properties of $g_{xc}$ and 
$\overline{g}_{xc}$ for the 3D uniform electron gas.
Equation~(\ref{eq_defg}) implies the positivity constraint
$g_{xc}\ge 0$ and the particle-conservation
sum rule, which can be divided into exchange and correlation,
\bea
\int_0^{\infty} du\,4\pi u^2 n\,(g_x-1) & = & -1 \label{eq_sumex}\\
\int_0^{\infty} du \,4 \pi u^2 n \,g_c = \int_0^{\infty} du \,4 \pi u^2 n \,
\gcav & = & 0.
\eea  
With the Coulomb interaction $1/u$,
the exchange function $g_x$, the correlation function $g_c$, and its 
coupling-constant averaged $\gcav$ 
integrate to the exchange energy $\epsilon_x$, to the
potential energy of correlation $v_c$, and to the 
correlation energy $\ec$ respectively,
\bea
\frac{1}{2}\int_0^{\infty}du\,4\pi u^2\frac{1}{u}n\,(g_x-1) & = &
\epsilon_x(r_s,\zeta), \\
\frac{1}{2}\int_0^{\infty}du\,4\pi u^2\frac{1}{u}n\, g_c & = &
v_c(r_s,\zeta) \label{eq_sumpot}\\
\frac{1}{2}\int_0^{\infty}du\,4\pi u^2\frac{1}{u}n\,\gcav & = &
\ec(r_s,\zeta). 
\label{eq_sumcorr}
\eea
For further discussion of the exchange hole density
$n\,(g_x-1)$ surrounding an electron, the correlation hole density $n\,g_c$, 
and the generalization of Eqs.~(\ref{eq_sumex})-(\ref{eq_sumcorr}) to 
non-uniform densities, see Refs.~\onlinecite{WDA} and~\onlinecite{LP}.
\par
The short-range behavior of $g_{xc}$ is determined by the $1/u$ Coulomb
repulsion, which gives rise to the cusp condition\cite{Kimball2}
\be
\frac{dg_{xc}}{du}\Big|_{u=0}=g_{xc}\Big|_{u=0}.
\label{eq_cusp}
\end{equation}
The function $\overline{g}_{xc}$ satisfies a modified cusp 
condition\cite{PW,GP} which can be derived from Eqs.~(\ref{eq_gavdef})
and~(\ref{eq_cusp}).
A quite accurate estimate of the $r_s$ and $\zeta$ dependence
of the short-range expansion coefficients of $g_{xc}$ 
and $\overline{g}_{xc}$ has been
recently obtained by solving a scattering problem in a
screened Coulomb potential which describes the effective electron-electron
interaction in a uniform electron gas -- 
the extended solution\cite{GP} of the Overhauser model.\cite{Ov}
(Classical electrons
at zero temperature would have $g_{xc}|_{u=0} = 0$, but nonzero values have a
nondivergent potential-energy cost according to Eq.~(\ref{eq_sumpot}) 
and for quantum
mechanical electrons lower the kinetic energy associated with the swerving
motion needed to keep two electrons from colliding.  Thus the right-hand
side of Eq.~(\ref{eq_cusp}) is nonzero, except in the low-density limit.
It is similarly nonzero for a gas of classical electrons at an elevated
temperature.\cite{DP})
\par
The long-range part of the nonoscillatory $g_{xc}$ corresponds 
to the small-$k$ behavior of the static structure factor, which
is determined by the plasmon contribution, proportional to $k^2$, and by
the single-pair and multipair quasiparticle-quasihole excitation contributions,
proportional to $k^5$ and $k^4$ respectively,\cite{NP58,pines}
\be
S_{xc}(r_s,\zeta,k\to 0)  =  \frac{k^2}{2\omega_p(r_s)} + O(k^4),
\label{eq_plas}
\end{equation}
where $\omega_p(r_s)=\sqrt{3/r_s^3}$ is the plasma frequency. 
Equation~(\ref{eq_plas}) is called the plasmon sum rule. There is no
$k^3$ term in the small-$k$ expansion\cite{iwa} of $S_{xc}$. 
Since, when $k\to 0$, the exchange-only static structure factor $S_x$
is equal to
\be
S_x(\zeta,k\to 0) = \frac{3}{8}\left[(1+\zeta)^{2/3}+(1-\zeta)^{2/3}\right]
\frac{k}{k_F}-\frac{k^3}{16 k_F^3},
\end{equation} 
there must be a linear term and a cubic term in the 
small-$k$ expansion of the correlation 
static structure factor $S_c$
which cancel with the exchange. In real space, these terms correspond
to long-range tails $\propto u^{-4}$ and $\propto u^{-6}$ 
respectively.\cite{PW,GSB2} The 
nonoscillatory exchange-correlation pair-distribution
function has a long-range tail\cite{GSB,GSB2} $\propto u^{-8}$.
As for more general densities, the exchange-correlation hole is more
localized around its electron than the exchange hole (and thus better
described by local or semi-local approximations for non-uniform
densities).
The high-density limit of the random-phase 
approximation (RPA) exactly describes\cite{WP} the nonoscillatory
long-range part of $g_{xc}$, recovering Eq.~(\ref{eq_plas}) through order
$k^2$. The absence of the $k^3$ term in the small-$k$ expansion of
$S_{xc}$ was demonstrated for the $\zeta=0$ gas by using exact 
frequency-moment sum rules.\cite{iwa} 
The same arguments should hold for the $\zeta\ne 0$ gas. Notice that
the cancellation of the $k^3$ terms is obtained from beyond-RPA
considerations.\cite{iwa}

    Armed with these exact constraints, we can discuss the strengths and
weaknesses of previous analytic models, which unlike our present model
break down\cite{Rassolov,comment} outside the metallic density range 
$1\lesssim r_s\lesssim 10$.

    The Perdew-Wang model\cite{PW} was largely based on first principles, plus
limited fitting to Quantum Monte Carlo data.  This model introduced the
high-density limit of the RPA as the long-range component of $g_{xc}$.  But
that limit was modelled crudely, leading to violation of the
particle-conservation sum rule (and thus to failure for 
$r_s\lesssim 0.1$).  The
model did not incorporate the plasmon sum rule, and produced an incorrect
$u^{-5}$ nonoscillatory long-range limit 
for $g_{xc}$.  The positivity constraint was violated at
low densities, a problem evaded by switching over to a different analytic
form for $r_s>10$.  In this model, the spin resolution of $g_{xc}$, 
even in its revised form,\cite{comment} is less reliable 
than the total $g_{xc}$.

     The model of Gori-Giorgi, Sacchetti, and Bachelet\cite{GSB} was based upon
extensive fitting to spin-resolved Quantum Monte Carlo data for $\zeta=0$,
and did not address nonzero $\zeta$.  Their model for $g_{xc}$,
unlike that of Perdew and Wang, was analytically Fourier-transformable to
$S_{xc}$. It incorporated the
particle-conservation and plasmon sum rules, and the correct $u^{-8}$
long-range limit for $g_{xc}$, but did not build in the important high-density
limit of the RPA for large $u$, leading to failure for $r_s\ll 0.8$.  Moreover,
small-$u$ errors of the Monte Carlo data were transferred into the 
model.\cite{GP}

\section{Nonoscillatory exchange hole}
\label{sec_ex}
We present here our nonoscillatory model for the exchange hole. This new model
satisfies exact short-range and long-range conditions up to the same order as
our correlation-hole model (Sec.~\ref{sec_corrhole}) does.\par
The exact exchange-only pair-distribution function for the uniform gas is
\bea
g_x(\zeta,k_Fu) & = & 1+\tfrac{1}{2}\{(1+\zeta)^2J[(1+\zeta)^{1/3}k_Fu] 
\nonumber \\
& & +(1-\zeta)^2J[(1-\zeta)^{1/3}k_Fu]\},
\eea
where
\be
J(y)=-\frac{9}{2}\left(\frac{\sin y-y \cos y}{y^3}\right)^2.
\end{equation}
Our nonoscillatory $\langle J(y)\rangle$ is parametrized as
\bea
\langle J(y)\rangle & = & \frac{-9}{4 y^4}\Big[1-e^{-A_xy^2}
\Big(1+A_x y^2+\frac{A_x^2 y^4}{2}+ \frac{A_x^3y^6}{3!}\Big)\Big]\nonumber \\
& &
 +e^{-D_x y^2}(B_x+C_x y^2+E_x y^4 +F_x y^6).
\label{eq_J}
\eea
This model is similar in spirit, but not in detail, to
those of Refs.~\onlinecite{PW} and~\onlinecite{EP}.  
The first term of Eq.~(\ref{eq_J}) achieves the
correct average long-range behavior $-\frac{9}{4}y^{-4}$ as $y\to\infty$, 
and is damped out at small $y$ by the first square bracket which varies from 
$y^8$ as $y\to 0$ to 1 as $y\to\infty$.  The second term then builds 
in the correct
small-$y$ behavior.  The Gaussians smoothly blend the two terms, but are not
motivated by any physical model.  The analytic forms and linear parameters
in Eq.~(\ref{eq_J}) are convenient for constraint satisfaction. The
separation into long-range and short-range parts, although somewhat
arbitrary, could be useful for the construction of new density
functionals. The spherical Fourier transform of $\langle J(y)\rangle$,
\be
\tilde{J}(k)=\int_0^{\infty}\langle J(y)\rangle y^2\frac{\sin(ky)}{ky}\,dy,
\label{eq_Jk}
\end{equation}
is also analytic and is reported in Appendix~\ref{app_exck}. 
The large-$y$ expansion of Eq.~(\ref{eq_J}) is
\be
\langle J(y\to \infty)\rangle = -\tfrac{9}{4}y^{-4}+O(e^{-y^2}),
\end{equation}
while the nonoscillatory average of the exact $J(y)$ also contains a 
$-\frac{9}{4}y^{-6}$ term (and no other long-range term). 
Such a term was included in the models of Refs.~\onlinecite{PW}
and~\onlinecite{EP}, but with a coefficient wrong in both sign and magnitude.
As explained in Sec.~\ref{sec_prop}, the exact nonoscillatory 
correlation hole has
long-range terms $y^{-4}$ and $y^{-6}$ which exactly cancel with the
exchange,\cite{GSB,GSB2,iwa} so that the exact 
nonoscillatory exchange-correlation hole 
has a long-range tail\cite{GSB,GSB2} $\propto u^{-8}$ which is
purely correlation. However, as
detailed in Sec.~\ref{sec_LR}, our nonoscillatory correlation-hole model
is built without a $u^{-6}$ long-range term, since this choice preserves 
a simple and useful scaling. We have thus also set the $y^{-6}$ term
to zero in our nonoscillatory exchange-hole model, in order to
have an exchange-correlation hole with the exact $u^{-8}$ long-range
behavior.

The six parameters  $A_x$ through $F_x$ are fixed by requiring
that (i) the particle-conservation
sum rule is fulfilled, (ii) our $g_x$ gives zero contribution to the
plasmon sum rule, (iii) our $g_x$ recovers the exact exchange energy,
(iv) our $g_x$ is exact at $u=0$ in obedience to the Pauli principle
in real space (two electrons of parallel spin cannot come together, since the
antisymmetry of the wavefunction makes this probability vanish), 
(v) our $g_x$ has the exact
second derivative at $u=0$, and (vi) the 
information entropy ${\cal S}[-J(y)]$,
\be
{\cal S}[-J(y)]=\int_0^{\infty}dy\,4\pi y^2J(y)\ln[-J(y)],
\label{eq_infS}
\end{equation}
is maximized.\cite{EP,AZA}
${\cal S}$ of Eq.~(\ref{eq_infS}) is not a
thermodynamic entropy but a mathematical one whose maximization ensures
that the analytic $J(y)$ has no structure beyond that imposed by the exact
constraints used to construct it.
The parameter values are $A_x=0.77$, $B_x=-0.5$, $C_x=-0.08016859$,
$D_x=0.3603372$, $E_x=0.009289483$, and $F_x=-0.0001814552$.

\begin{figure}
\includegraphics[width=5.5cm]{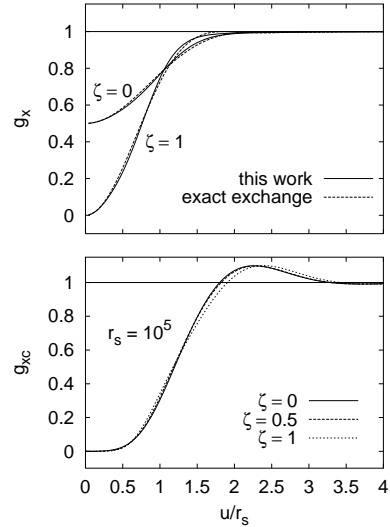} 
\caption{Upper panel: our nonsoscillatory model for 
exchange in the uniform electron gas
is compared with the exact Hartree-Fock curve.
Note that $g_x$ is the $r_s\to 0$ limit of $g_{xc}$.
Lower panel: low-density limit of our analytic model for the
exchange-correlation pair-distribution function of the uniform gas.
In this limit, the
model $g_{xc}$ is almost exactly independent of the relative spin polarization
$\zeta$.}
\label{fig_LD}
\end{figure}

Our nonoscillatory model $g_x$ is compared with the exact exchange at
$\zeta=0$ and $\zeta=1$ in the upper panel of Fig.~\ref{fig_LD}.
In the first panel of Fig.~\ref{fig_gLR}, the exchange hole
$g_x-1$ is multiplied by $(u/r_s)^4$ in order to show how
our model (solid line) averages out the oscillations of
the exact exchange hole (dashed line).
\section{Nonoscillatory correlation hole}
\label{sec_corrhole}
Following Perdew and Wang,\cite{PW} we write the
nonoscillatory part of the correlation hole as
the sum of a long-range part and a short-range part,
somewhat as in Eq.~(\ref{eq_J}):
\bea
\langle \overline{g}_c(r_s,\zeta,k_F u)\rangle & & = 
\frac{\phi^3r_s}{\kappa}
\frac{\overline{f}_1(v)}{(k_Fu)^2}\,\Big[1-e^{-d\,x^2}\Big(1+d\,x^2
 \nonumber \\
& & 
+\frac{d^2}{2}\,x^4\Big)\Big]
+e^{-d\,x^2}\sum_{n=1}^6 c_n\, x^{n-1},
\label{eq_base}
\eea
where $\kappa=(4/3\pi)(9\pi/4)^{1/3}$, $\phi=[(1+\zeta)^{2/3}+
(1-\zeta)^{2/3}]/2$, $x=k_Fu/\phi$,
and $v=\phi\kappa\sqrt{r_s}k_Fu$. The six linear parameters
$c_n$ depend on both $r_s$ and $\zeta$, while the nonlinear
parameter $d$ only depends on $\zeta$.

The first term in the r.h.s.~of Eq.~(\ref{eq_base}) is 
the long-range part of our $\gcav$: the function
$\overline{f}_1(v)$ is a new parametrization (see Sec.~\ref{sec_LR})
of the RPA limit found by Wang and Perdew\cite{WP} 
and displayed in Fig.~2 of Ref.~\onlinecite{PW}. We multiplied
$\overline{f}_1(v)/(k_Fu)^2$ by a cutoff function which cancels
its small-$u$ contributions, so that the long-range
part of our $\gcav$ vanishes through order $u^4$ and does not interfere
with the short-range part.
\par
\begin{figure}
\includegraphics[width=7cm]{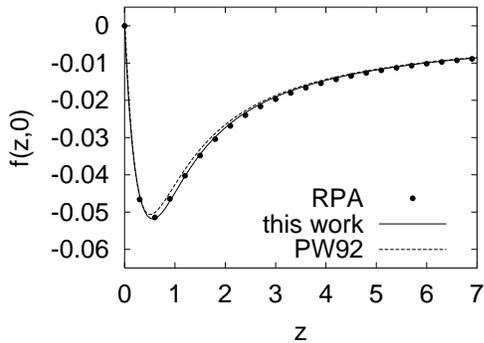} 
\caption{The function $f(z,0)$ given in Ref.~\protect\onlinecite{WP}.
The exact calculation (RPA) is compared with the present parametrization and
with the one of Perdew and Wang\cite{PW} (PW92).}
\label{fig_fz}
\end{figure}
For modeling 
the short-range part, corresponding to the last term in the 
r.h.s.~of Eq.~(\ref{eq_base}), we use our recent results obtained
by solving the Overhauser model,\cite{GP} which allow us to fix
the $r_s$ and $\zeta$ dependence of the linear parameters 
$c_1$, $c_2$ and $c_3$ (Sec.~\ref{sec_SR}).  
We then use the remaining three linear parameters,
$c_4$, $c_5$ and $c_6$, to fulfill the particle-conservation sum rule
and the plasmon sum rule, and to recover the ``exact'' correlation energy
(Sec.~\ref{sec_sumrules}).
Finally, the nonlinear parameter $d(\zeta)$, which determines the ``mixing'' of
long-range and short-range contributions, is fixed by imposing the
positivity constraint on $g_{xc}$ when $r_s \to \infty$ (Sec.~\ref{sec_d}).
\begin{figure}
\includegraphics[width=\columnwidth]{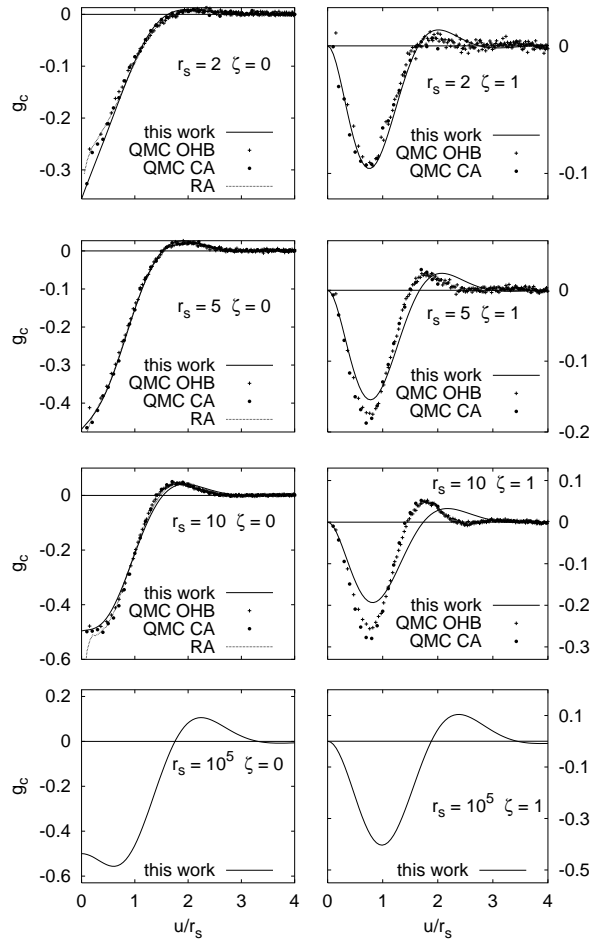} 
\caption{Coulomb correlation contribution $g_c$
to the pair-distribution function $g_{xc}$ for the uniform electron gas
for the paramagnetic ($\zeta=0$) and ferromagnetic ($\zeta=1$) state.
Our new analytic model is compared with the Diffusion Quantum Monte Carlo
results of Ortiz, Harris, and Ballone\cite{OHB} (OHB), and of
Ceperley and Alder\cite{cepald} (CA). The pair-correlation function
corresponding to the local-field-factor model of Richardson and
Ashcroft\cite{RA} (RA) is also shown. In the two bottom panels, the 
low-density limit of our $g_c$ is reported.}
\label{fig_gcQMC}
\end{figure}
\subsection{Long-range part}
\label{sec_LR}
As discussed in Refs.~\onlinecite{WP} and~\onlinecite{PW},
the long-range ($u\to \infty$) part of the nonoscillatory
correlation hole can be obtained from the
random-phase approximation by computing its $r_s\to 0$ 
limit. One finds
\be
n\langle\gcav(r_s,\zeta,k_Fu)\rangle\to \phi^3(\phi k_s)^2
\frac{\overline{f}_1(v)}{4\pi v^2},
\label{eq_lr}
\end{equation}
where $k_s$ is the Thomas-Fermi screening wave vector, $k_s=
\kappa \sqrt{r_s}k_F$. The function $\overline{f}_1(v)$ is
the spherical Fourier transform of the function $f(z,0)$
given by Eqs.~(29), (34) and~(36) of Ref.~\onlinecite{WP},
\be
\overline{f}_1(v)=2v^2\int_0^{\infty}dz\,z^2\,f(z,0)\,\frac{\sin(vz)}{vz},
\end{equation}
where $z=k/\phi k_s$ is the
proper scaled variable in reciprocal space.
The small- and large-$z$ expansion
of $f(z,0)$ is
\bea
f(z\to 0,0) & = & -\tfrac{3}{\pi^2}z+\tfrac{4\sqrt{3}}{\pi^2}z^2+O(z^3)
\label{eq_fzsmallz} \\
f(z\to \infty,0) & = & -\tfrac{2(1-\ln 2)}{\pi^2}z^{-1}+O(z^{-2}).
\label{eq_fzlargez}
\eea
Equation~(\ref{eq_fzlargez}) gives the high-density limit of
the corresponding correlation energy,
\be
\ec(r_s\to 0,\zeta) =\tfrac{(1-\ln 2)}{\pi^2} \phi(\zeta)^3 \ln r_s+O(r_s^0),
\label{eq_ecHD}
\end{equation}
which is exact at $\zeta=0$ and 1, but is slightly different from the
exact result for $0<\zeta<1$ (see Refs.~\onlinecite{PW} and~\onlinecite{WP}
for further details).
The small-$z$ expansion of $f(z,0)$,
Eq.~(\ref{eq_fzsmallz}), fulfills the particle-conservation
sum rule [$f(z=0,0)=0$], contains a linear term which cancels
with the exchange (and corresponds to a long-range
tail $\propto u^{-4}$ in real space, see Sec.~\ref{sec_prop}), and fulfills 
the plasmon sum rule [exact $z^2$ coefficient, see Eq.~(\ref{eq_plas})].
The $z^3$ term in Eq.~(\ref{eq_fzsmallz}), if it does not vanish, 
produces a $u^{-6}$ contribution to the correlation hole at large $u$.
\par
\begin{figure}
\includegraphics[width=6.7cm]{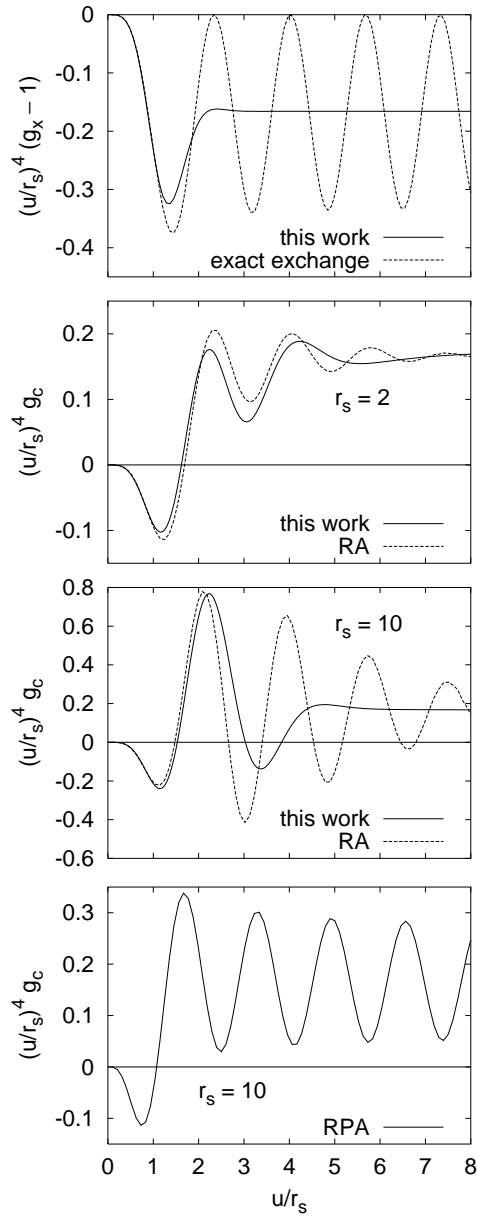} 
\caption{Upper panel: long-range part of the exchange hole.
Our nonsoscillatory model is compared with the exact
exchange. Second and third panel: long-range part of the correlation
hole. Our nonoscillatory model is compared with $g_c$ obtained from
the Richardson and Ashcroft\cite{RA} (RA) local-field factor. In the
lowest panel the random-phase-approximation (RPA) result for $r_s=10$
is also shown. All curves are for the $\zeta=0$ gas.}
\label{fig_gLR}
\end{figure}

 
As said in Secs.~\ref{sec_prop} and~\ref{sec_ex}, the long-range 
($u\to\infty$)
nonoscillatory behavior of the exact exchange hole contains $u^{-4}$ 
and $u^{-6}$ contributions which are cancelled\cite{GSB,GSB2,iwa} by 
similar contributions to the exact correlation hole.  
When we use the high-density limit of Eq.~(\ref{eq_lr})
for the long-range part of the correlation hole, we automatically achieve
cancellation of the $u^{-4}$ terms.  But to cancel the $u^{-6}$ terms 
in $g_x-1$, we would have to replace $\overline{f}_1(v)/v^2$ in 
Eq.~(\ref{eq_lr}) by $\overline{f}_1(v)/v^2 +
r_s \phi h(r_s,\zeta,v)$, where $\overline{f}_1(v)/v^2$ has no 
$v^{-6}$ contribution and $h$ is proportional to $v^{-6}$ 
with no $r_s$ or $\zeta$ dependence at large $v$.  
The extra term $r_s\phi h$ vanishes
in the high-density limit for a given $v$, and is unknown.  
Since we want to keep for our
$\gcav$ the simple form of Eq.~(\ref{eq_base}), but we also want to have 
the correct long-range behavior ($\propto u^{-8}$)
for $\gxcav$, we decided simply to set the
$u^{-6}$ terms to zero in both our exchange (Sec.~\ref{sec_ex}) 
and correlation-hole models.  
Figures~\ref{fig_gcQMC} and~\ref{fig_gLR}
do not suggest that this choice introduces any
significant error into our models for the separate exchange and
correlation holes.

We thus parametrize $\overline{f}_1(v)$ as follows
\be
\overline{f}_1(v)=\frac{a_0+b_2 v+a_1 v^2+a_2 v^4+a_3 v^6}{(v^2+b^2)^4}.
\label{eq_newf1}
\end{equation}
With respect to the parametrization given by Perdew and Wang,\cite{PW} our
Eq.~(\ref{eq_newf1}) has the advantage that it is analytically
Fourier-transformable (see Appendix~\ref{app_fz}), so that
the particle-conservation sum rule and the plasmon sum rule can
be easily imposed. (They are not fulfilled by the Perdew and 
Wang\cite{PW} parametrization). After imposing on our $\overline{f}_1(v)$
all the exact properties plus the vanishing of the $z^3$ term
in Eq.~(\ref{eq_fzsmallz}),
we are left with one free parameter, $b$, which is
fixed by a best fit to our RPA data.\cite{WP} All the parameter values
are reported in Appendix~\ref{app_fz}. The function $f(z,0)$ corresponding
to our parametrization [see Eq.~(\ref{eq_SFTf1})] is compared 
in Fig.~\ref{fig_fz} with the
RPA result and with the Fourier transform of the Perdew and
Wang\cite{PW} (PW92) $\overline{f}_1(v)$.
\subsection{Short-range part}
\label{sec_SR}
Our $\gcav$ has the small-$u$ expansion
\be
\langle\gcav\rangle=c_1+c_2\kfr+(-c_1d+c_3)\left(\kfr\right)^2+
O(u^3).
\end{equation}
In order to recover the short-range behavior
obtained by solving the Overhauser model,\cite{GP} we require
\bea
c_1 & = & \tfrac{(1-\zeta^2)}{2}\left[\overline{a}_0^{\ud}(r_s^{\ud})-1\right]
\label{eq_c1}\\
c_2 & = & \phi\left(\tfrac{4}{9\pi}\right)^{1/3}\tfrac{
(1-\zeta^2)}{2}\tfrac{[(1+\zeta)^{1/3}+(1-\zeta)^{1/3}]}{2}\overline{a}_1^{\ud}
(r_s^{\ud})
\label{eq_c2} \\
c_3 & = & \phi^2\left[\left(\tfrac{4}{9\pi}\right)^{2/3}\overline{a}_2(r_s,\zeta)
-\tfrac{(1+\zeta)^{8/3}+(1-\zeta)^{8/3}}{20}\right]+ \nonumber \\
& & +c_1d,\label{eq_c3}
\eea
where $r_s^{\ud}=2r_s/[(1+\zeta)^{1/3}+(1-\zeta)^{1/3}]$, and 
$\overline{a}_0^{\ud}$,
$\overline{a}_1^{\ud}$ and $\overline{a}_2(r_s,\zeta)$ are given by 
Eqs.~(36), (37) and (46) of Ref.~\onlinecite{GP}. In this way, the modified
cusp condition is exactly satisfied.\cite{GP}
\subsection{Sum rules}
\label{sec_sumrules}
We want our correlation hole to satisfy the particle-conservation
sum rule and the plasmon sum rule, and to recover 
the ``exact'' correlation energy.
Our new parametrization of the function $\overline{f}_1(v)$ 
satisfies the particle-conservation sum rule, and recovers the exact 
plasmon coefficient and the $\ln r_s$ term of the resulting correlation
energy. Thus, we only have to require that the remaining part of our $\gcav$ 
gives zero contribution to (i) the particle-conservation sum rule and (ii)
the plasmon sum rule, and (iii) recovers the correlation energy beyond the 
$\ln r_s$ term. 
In this way we have three linear equations for the three parameters $c_4$,
$c_5$ and $c_6$:
\bea
& & \sum_{n=1}^6 \tilde{c}_n\int_0^{\infty}e^{-t^2}t^{n+1}dt  = 
A\,S(\alpha) \\
& & \sum_{n=1}^6 \tilde{c}_n\int_0^{\infty}e^{-t^2}t^{n+3}dt  = 
A\,P(\alpha) \\
& & \sum_{n=1}^6 \tilde{c}_n\int_0^{\infty}e^{-t^2}t^ndt  = 
-A\,R(\alpha)+E,
\eea
where $\tilde{c}_n=c_n/d^{\frac{n-1}{2}}$, $t=\sqrt{d}k_Fu/\phi$, $A=\phi r_s 
d/\kappa$, $\alpha=\phi^2\kappa (r_s/d)^{1/2}$, and
\bea
 S(\alpha)  =  \int_0^{\infty}\overline{f}_1(\alpha t)
e^{-t^2}\left(1+t^2+\tfrac{1}{2}t^4\right)dt 
\label{eq_Sa}\\
 P(\alpha)  =  \int_0^{\infty}\overline{f}_1(\alpha t)
e^{-t^2}t^2\left(1+t^2+\tfrac{1}{2}t^4\right)dt 
\label{eq_Pa}\\
 R(\alpha)  =  \int_0^{\infty}\frac{\overline{f}_1(\alpha t)}{t}
\left[1-e^{-t^2}\left(1+t^2+\tfrac{1}{2}t^4\right)\right]dt 
\label{eq_Ra}\\
 E = \frac{2r_s d}{3\,\phi^2}
\left(\frac{9\pi}{4}\right)^{2/3}\epsilon_c(r_s,\zeta).
\label{eq_EE}
\eea
The functions $S(\alpha)$, $P(\alpha)$ and $R(\alpha)$ can be
obtained analytically and are reported in Appendix~\ref{app_SPR}.
The parameters $c_4$, $c_5$ and $c_6$ are then equal to
\bea
\tilde{c}_4 & = & \{100\sqrt{\pi}(3\pi-8)\tilde{c}_1+
(690\pi-2048)\tilde{c}_2+\sqrt{\pi}(225\pi
 \nonumber \\
& & -672)\tilde{c}_3+(8192-2100\pi)AS(\alpha)+AP(\alpha)(600\pi-
\nonumber \\
& & 2048)+960\sqrt{\pi}[AR(\alpha)-E]\}/[4(512-165\pi)]
\label{eq_c4}\\
\tilde{c}_5 & = & 2\{(30\pi-128)\tilde{c}_1-
8\sqrt{\pi}\tilde{c}_2+(39\pi-128)\tilde{c}_3-
 \nonumber \\
& & 144\sqrt{\pi}AS(\alpha)+16\sqrt{\pi}AP(\alpha)-256[AR(\alpha)
\nonumber \\
& & -E]\}/(512-165\pi)\label{eq_c5}\\
\tilde{c}_6 & = & \{\sqrt{\pi}(180\pi-624)\tilde{c}_1+
(150\pi-512)\tilde{c}_2+\sqrt{\pi}(135\pi
 \nonumber \\
& & -432)\tilde{c}_3+(3072-1260\pi)AS(\alpha)+AP(\alpha)(360\pi-
\nonumber \\
& & 1024)-480\sqrt{\pi}[AR(\alpha)-E]\}/[6(165\pi-512)].
\label{eq_c6}
\eea
\subsection{Positivity constraint in the low-density limit}
\label{sec_d}
The nonlinear parameter $d$ can be fixed by imposing the condition that
$\overline{g}_{xc}$ remains positive when $r_s\to\infty$.
The short-range behavior imposed on our $\gcav$ 
ensures that the small-$u$ expansion of the corresponding
$\overline{g}_{xc}$ has coefficients which
are always $\ge 0$ through order $u^2$, and which become
zero in the low-density or strongly-correlated limit. We have checked that,
if we want
to have a positive $\overline{g}_{xc}$ for all densities, we only need
to require that also the $u^3$ coefficient (equal to $c_4-d\,c_2$)
becomes
0 when $r_s\to\infty$, according to the cusp condition for parallel-spin 
pairs.\cite{Kimball2,GSB,GP} We thus have an equation for $d(\zeta)$:
\be
\lim_{r_s\to \infty} c_4(r_s,\zeta)-d(\zeta)c_2(r_s,\zeta)=0.
\label{eq_dzld}
\end{equation}
Equation~(\ref{eq_dzld}) is rather complicated since $c_4$ also depends
nonlinearly on $d$. However, it
can be solved numerically for each $\zeta$, and,
when the Perdew-Wang\cite{PWEc} parametrization of
the correlation energy is used in Eq.~(\ref{eq_EE}), 
the result is very well fitted by
\be
d(\zeta)=d(0)\left[(1+\zeta)^{2/3}+(1-\zeta)^{2/3}-1\right],
\label{eq_dz}
\end{equation} 
with $d(0)=0.131707$.
\begin{figure}
\includegraphics[width=7cm]{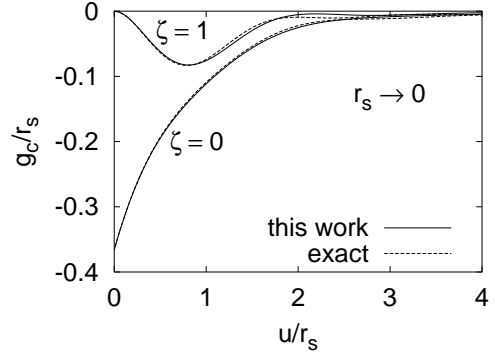} 
\caption{Coulomb correlation contribution to the
pair-distribution function for the uniform electron gas
for the paramagnetic ($\zeta=0$) and ferromagnetic ($\zeta=1$) state
in the high density ($r_s\to 0$) limit. The result from our analytic
model is compared with the exact calculation of 
Refs.~\protect\onlinecite{Rassolov,reply}.}
\label{fig_gcHD}
\end{figure}
\section{Results for the exchange-correlation hole}
\label{sec_res}
In the next three subsections we present and discuss our results
for the nonoscillatory $g_x$, $g_c$ and $g_{xc}$ in the whole
($0< r_s < \infty$) density range.
We have used the correlation energy $\ec$ as parametrized by Perdew
and Wang,\cite{PWEc} which was built with the Quantum Monte Carlo data
of Ref.~\onlinecite{cepald} as an input. It is however straightforward to
build into our equations an {\it ab initio} $\ec$ for the 3D uniform
gas when available,\cite{mike} showing that the exact constraints
suffice to determine $g_{xc}$ without the need for any 
``numerical experiment''.
\subsection{Metallic densities}
In the six upper panels of Fig.~\ref{fig_gcQMC} we compare
our analytic $g_c$ with the Quantum Monte Carlo (QMC) data
of Ceperley and Alder\cite{cepald} (CA) and of Ortiz, Harris and
Ballone\cite{OHB} (OHB) for $r_s =2$, 5 and 10, and for $\zeta=0$
(left) and $\zeta=1$ (right). In the $\zeta=0$ case, we also report
$g_c$ as obtained by the dynamic local-field-factor model
of Richardson and Ashcroft\cite{RA} (RA) via the fluctuation-dissipation 
theorem (as in Ref.~\onlinecite{LGP}). The RA model yields very
accurate correlation energies $\ec(r_s,\zeta=0)$,\cite{LGP} and
we find that the RA $g_c$ is in very good agreement with QMC data except at
small $u$. The limit $u\to 0$ is not
correctly included in the RA parametrization of the local-field factor,
which violates the Pauli principle in real space.

We see that our model is in fair agreement with QMC data
for the paramagnetic gas. In the ferromagnetic case, where the pair-correlation
function shows stronger oscillations even at intermediate densities, the
agreement is less satisfactory (as in the model of Ref.~\onlinecite{PW}). 
This is not surprising, since
our model does not take into account the
energetically unimportant oscillations: it only includes the minimum number
of oscillations needed to fulfill the sum rules. This
is evident in the second and third panel of Fig.~\ref{fig_gLR}, where
$g_c$ is multiplied by $(u/r_s)^4$. In this way, the long-range
oscillations are amplified and become clearly visible even at 
metallic densities:
we can thus compare our model (solid line) with the RA result (dashed line). 
This is done at $r_s=2$ and 10. The many exact properties 
imposed on the RA local-field factor and the first three left panels of 
Fig.~\ref{fig_gcQMC} suggest that the long-range part of the RA $g_c$ is
very reliable and that the oscillations are probably accurately 
described. One clearly sees in Fig.~\ref{fig_gLR} how our model 
follows the first oscillation and averages
out the others. In the lowest panel of Fig.~\ref{fig_gLR},
the long-range oscillations of the random-phase approximation (RPA) $g_c$ at 
$r_s=10$ are also shown. At large $r_s$, the RPA oscillations of $g_c$ tend
to cancel the ones of $g_x$ (first panel), while the effect of a 
dynamic local-field factor clearly inverts this tendency: the oscillations
of the RA $g_c$ (second and third panel) are almost in phase with the 
oscillations of $g_x$. We interpret this to
mean that the RA $g_{xc}$ of the low-density uniform electron gas is building
up an incipient Wigner-crystal-like order of the other electrons around a
given electron. 
\subsection{High density}
In the high-density limit, $g_c=2\gcav$ goes to zero, so 
that $g_{xc}\to g_x$. 
It has been shown\cite{PW,Rassolov,reply} that
in the $r_s\to 0$ limit $g_c/r_s$ remains finite and goes to a well defined
function of $u/r_s$, which has been computed exactly.\cite{Rassolov,reply}
In Fig.~\ref{fig_gcHD} we compare this exact calculation (dashed line)
with our model (solid line), computed at $r_s=10^{-5}$, 
for $\zeta=0$ and $\zeta=1$.
We see that (i) our model does not break down as $r_s\to 0$, and (ii)
there is fair agreement with the exact result. Previous 
models\cite{PW,GSB} for $g_c$ usually break down at $r_s\sim 0.1$.
Feature (i) is due to the new parametrization of $\overline{f}_1(v)$ which
exactly fulfills the particle-conservation
sum rule, while feature (ii) is due to the
short-range behavior taken from Ref.~\onlinecite{GP}, which includes the exact
high-density limit of the short-range coefficients.
\begin{figure}
\includegraphics[width=6.7cm]{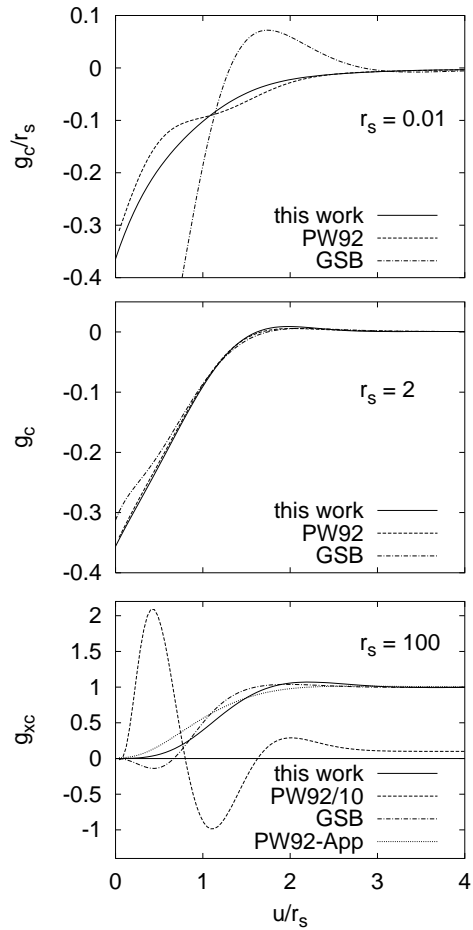} 
\caption{Comparison of the present work with the models of 
Ref.~\protect\onlinecite{PW} (PW92) and of Ref.~\protect\onlinecite{GSB}
(GSB)
at high densities (first panel), metallic densities (second panel) and
in the low-density regime (third panel). In the $r_s=100$ case the original
PW92 curve has been divided by 10, and the low-density form proposed
in the Appendix of Ref.~\protect\onlinecite{PW} (PW92-App) is also reported.
All curves are for the paramagnetic ($\zeta=0$) gas.}
\label{fig_gold}
\end{figure}
\begin{figure}
\includegraphics[width=6.7cm]{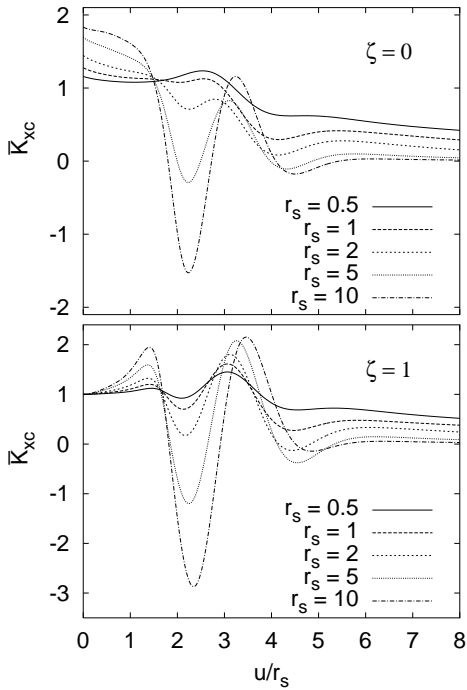} 
\caption{The correlation factor $\overline{K}_{xc}$
defined in Eq.~(\ref{eq_Kxc}) for
the paramagnetic (upper panel) and ferromagnetic (lower panel)
uniform electron gas.}
\label{fig_Kxc}
\end{figure}
\subsection{Low density}
In the low-density or strongly-correlated limit,
we expect that $g_{xc}$ (equal to $\overline{g}_{xc}$ in this case)
does not depend on $\zeta$, since in this 
limit the Pauli principle in real space becomes irrelevant
with respect to the Coulomb repulsion. In the lower panel
of Fig.~\ref{fig_LD} we report our model at $r_s=10^5$ for
three different values of the spin polarization $\zeta$. We see that
the $\zeta$ dependence of our low-density $g_{xc}$ is indeed very weak, 
and that, unlike previous parametrizations,\cite{PW,GSB}
our model never gives rise to an unphysical negative pair-distribution
function. Figure~\ref{fig_LD} also offers a view on the same
scale of the extreme high-density limit of $g_{xc}$ (equal to the
exchange-only pair-distribution function, first panel) and of the
extreme low-density limit (second panel). We see how the $\zeta$ dependence
of $g_{xc}$, which is very strong in the $r_s\to 0$ limit,
is cancelled by correlation in the $r_s\to \infty$ limit.  
The low-density limit of our $g_c=g_{xc}-g_x$ is reported in the
two lowest panels of Fig.~\ref{fig_gcQMC} for $\zeta=0$ or
$\zeta=1$.
\subsection{Comparison with previous analytic models}
In Fig.~\ref{fig_gold} the present model is compared with the
parametrizations of Perdew and Wang\cite{PW} (PW92) and of
Gori-Giorgi, Sacchetti and Bachelet\cite{GSB} (GSB). In the
first panel, we see that in the high-density regime ($r_s=0.01$)
the PW92 model starts to break down,\cite{comment,reply}
and that the GSB parametrization is completely unable to describe 
such high densities. (This is due to the wrong $r_s\to 0$ behavior of the 
GSB on-top pair density.)
At $r_s=2$, well inside the metallic regime, we see (second panel) that
the present work is very close to the PW92 model and slightly deviates from
the GSB curve at $u/r_s\lesssim 1$. Finally, in the third panel we show
the total pair-distribution function $g_{xc}$ at $r_s=100$: the PW92 model
in its original form completely blows up, while the GSB model becomes
negative at $u/r_s\lesssim 1$ but is still ``reasonable''. The low-density
form proposed in the Appendix of Ref.~\onlinecite{PW} (PW92-App) is
also reported: it corresponds to an exchange-correlation hole
narrower than the present one.
\subsection{Features of the ``correlation factor''}
To better see the effects of correlation, we define a ``correlation
factor''
\begin{equation}
\overline{K}_{xc}(r_s,\zeta,k_Fu)=\frac{\gxcav-1}{g_x-1}=
1+\frac{\gcav}{g_x-1}
\label{eq_Kxc}
\end{equation}
which morphs the exchange hole into the exchange-correlation hole, and
is displayed in Fig.~\ref{fig_Kxc}.  We must of course use 
non-oscillatory models here, since the exact $g_x-1$ has nodes which 
would create singularities in Eq.~(\ref{eq_Kxc}).  Figure~\ref{fig_Kxc} 
shows that $\overline{K}_{xc} \to K_x=1$ in the $r_s\to 0$ limit.
For typical valence-electron densities, we see that correlation enhances
or deepens the hole ($\overline{K}_{xc}>1$) around an electron for 
$u/r_s\lesssim 1.5$, while
it screens out the long-range part of the hole.  Because of the exact
cancellation of the $u^{-4}$ and $u^{-6}$ long-range terms between 
$\gcav$ and $g_x-1$, $\overline{K}_{xc}$ at large $u$ goes to 0 like 
$u^{-4}$.  For $r_s>2$, $\overline{K}_{xc}$ can
be negative in the range $1.5\lesssim u/r_s\lesssim 3$, corresponding to a 
positive peak in $\gxcav-1$.  We can think of 
$\overline{K}_{xc}(r_s,\zeta,k_Fu)/u$ as an
effective, density-dependent screened electron-electron interaction whose
exchange energy equals the exchange-correlation energy of the Coulomb
interaction $1/u$.
\par 
The correlation factor has a possible application\cite{BPE,tao} to the
modelling of exchange and correlation in systems of non-uniform density.
First we note that the exchange-correlation energy is fully determined by
the spherical average $n_{xc}(\rv,u)$ of the hole,\cite{Gunreview}
\begin{equation}
n_{xc}(\rv,u)=\int\frac{d\Omega_{\uv}}{4\pi}n_{xc}(\rv,\rv+\uv).
\end{equation}
A possible ``correlation factor model''\cite{tao} for $n_{xc}(\rv,u)$ is
\begin{equation}
n_{xc}(\rv,u)=\overline{K}_{xc}(\rv,u)\,n_{x}(\rv,u),
\label{eq_CFM}
\end{equation}
where $n_{x}(\rv,u)$ is the exact exchange hole. $\overline{K}_{xc}$ for a
non-uniform density could be constructed from Eq.~(\ref{eq_Kxc}) by inserting 
into Eq.~(\ref{eq_base}) (or a simplification thereof)
an $\rv$-dependent set of linear parameters 
$c_n$ chosen to satisfy exact constraints on $n_{xc}$.  
The result would presumably be a model
for exact exchange and approximate correlation compatible therewith.
The screening of the long-range part of the exact exchange hole is
essential for a proper description of molecules.\cite{HC}
\section{Spin resolution}
\label{sec_spinres}
We can define spin-resolved pair-distribution functions which
describe spatial correlations between $\uu$, $\dd$, and $\ud$ electron
pairs. Their normalization is such that the spin-averaged
$g_{xc}$ of Eq.~(\ref{eq_defg}) is equal to
\begin{equation}
g_{xc}=\left(\frac{1+\zeta}{2}\right)^2g_{xc}^{\uu}+
\left(\frac{1-\zeta}{2}\right)^2g_{xc}^{\dd}+
\left(\frac{1-\zeta^2}{2}\right)g_{xc}^{\ud}.
\end{equation}
While the spin resolution of the exchange-only pair-distribution
function $g_x$ is well known,\cite{PW} the correlation part
is much more delicate, and an accurate analytic representation
is only available\cite{GSB} for $\zeta=0$ in the density 
range $0.8\le r_s\le 10$. 

The model presented in Sec.~\ref{sec_corrhole} can be used
to build up spin-resolved correlation functions provided that
the spin resolution of the input quantities is known.
The input quantities are (i) the RPA long-range part, (ii)
the short-range coefficients from the solution of the Overhauser model, and
(iii) the correlation energy. 
Once these input quantities are known, in fact, one can build, say, 
$\overline{g}_c^{\ud}$, starting from the same Eq.~(\ref{eq_base}),
using the RPA $\ud$ long-range part, and putting the 
$\ud$ short-range coefficients into Eqs.~(\ref{eq_c1})-(\ref{eq_c3}),
and $\ec^{\ud}$ into Eq.~(\ref{eq_EE}). Finally,
the positivity constraint of $g_{xc}^{\ud}$ in the low-density limit 
can be applied to find $d_{\ud}(\zeta)$, as done in Sec.~\ref{sec_d}.

The first point is thus to see whether the quantities (i)-(iii) are
available in their spin-resolved contributions.
The RPA long-range part is easily spin-resolved
for the $\zeta=0$ gas,\cite{GSB,comment} while its spin resolution in the
partially polarized gas is less trivial. The short-range coefficients
from the Overhauser model are  available as $\uu$, $\dd$ and
$\ud$ separate contributions.\cite{GP} The correlation energy
represents the major problem: at $\zeta=0$ it can be easily spin 
resolved in the 
high- and low- density limits, while at intermediate densities 
the best estimate is probably the one given in Ref.~\onlinecite{GSB}.
Almost nothing about the spin resolution of $\ec$ is known for 
the $\zeta\ne 0$ gas, except in the extreme
low-density limit, when the system becomes $\zeta$-independent.   

Here, we show results for $g_c^{\ud}$ in three cases: 
the extreme low-density limit, the high-density limit of the
paramagnetic gas, and the $r_s=2,\; \zeta=0$ case.
The low-density limit must be treated first, since it is necessary
to determine $d_{\ud}(\zeta)$ through the positivity constraint
on $g_{xc}^{\ud}$ when $r_s\to\infty$.

\begin{figure}
\includegraphics[width=7cm]{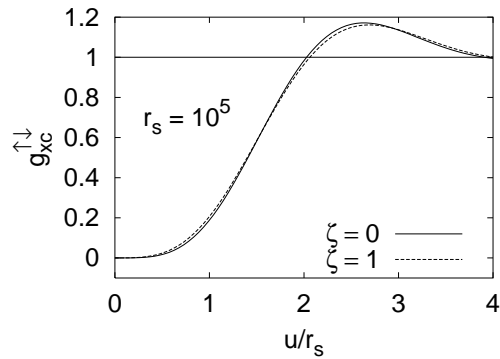} 
\caption{Up-down pair-distribution function for the 
uniform electron gas in the low-density limit.}
\label{fig_gudLD}
\end{figure}

When $\zeta=0$, the spin-resolution within RPA is very simple:
up-up and up-down interactions contribute the same amount to
correlation.\cite{GSB,comment} The long-range part of our 
$\overline{g}_c^{\ud}$ can thus be built using
the function $\overline{f}_1(v)$ of Eq.~(\ref{eq_newf1}) with the same
parameters of Appendix~\ref{app_fz}. While the 
spin-averaged nonoscillatory long-range
behavior computed within RPA is also exact beyond it at all densities,
its spin resolution is exact beyond RPA only when $r_s\to 0$.\cite{nota} 
We keep on using it even in the 
extreme low-density limit, since it is the only way to
build up a spin-resolved $g_c$ starting from our model. 
As we shall see, the results obtained are reasonable, and
justify our choice. 
When $r_s\to \infty$, we expect
the statistics to be energetically unimportant,\cite{nota2} so that
$\epsilon_{xc}^{\uu}=\epsilon_{xc}^{\dd}=\epsilon_{xc}^{\ud}
=\epsilon_{xc}$. We thus find $\ec^{\ud}=\epsilon_{xc}=
-0.892/r_s$, where the numerical coefficient corresponds to the
Perdew-Wang parametrization\cite{PWEc} of $\ec$. The
positivity constraint on $g_{xc}^{\ud}$ gives
\be
d_{\ud}(\zeta)=d_{\ud}(0)\left[(1+\zeta)^{2/3}+(1-\zeta)^{2/3}-1\right],
\end{equation}
with $d_{\ud}(0)=0.0885717$. The results for $g_{xc}^{\ud}$ are
shown in Fig.~\ref{fig_gudLD}, at $r_s=10^5$, for $\zeta=0$ and $\zeta=1$.

\begin{figure}
\includegraphics[width=7cm]{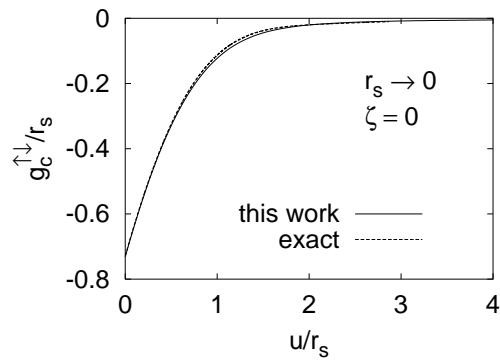} 
\caption{Up-down Coulomb correlation contribution to the
pair-distribution function for the 
paramagnetic ($\zeta=0$) uniform electron gas
in the high density ($r_s\to 0$) limit. The result from our analytic
model is compared with the exact calculation of 
Ref.~\protect\onlinecite{reply}.}
\label{fig_gcudHD}
\end{figure}

For the high-density limit of the paramagnetic gas,
all the spin-resolved input quantities are exactly known. It is thus
the best case to test our model. When $r_s\to 0$, the
spin-resolution from RPA is exact also beyond it: the long-range part
of $\overline{g}_c^{\ud}$ is in this case {\em exactly} described by
Eq.~(\ref{eq_newf1}) with the parameters of Appendix~\ref{app_fz}.
The correlation energy, in this limit,\cite{GSB,GSB2} is simply 
equal to the spin-averaged
correlation energy of Eq.~(\ref{eq_ecHD}) with $\zeta$ set to zero.
The short-range $\ud$ coefficients from Ref.~\onlinecite{GP} include
the exact spin-resolved high-density limit of the $\zeta=0$ gas. 
The so-obtained $g_c^{\ud}$ is shown in Fig.~\ref{fig_gcudHD},
together with the exact calculation from Ref.~\onlinecite{reply}, which
is, in this case, equal to the RPA result. We find very good agreement.

\begin{figure}
\includegraphics[width=7cm]{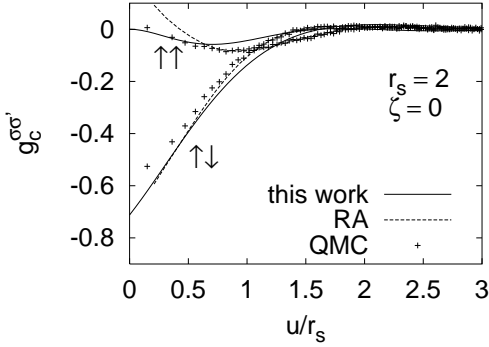} 
\caption{Spin-resolved Coulomb correlation contribution to the
pair-distribution functions for the 
paramagnetic uniform electron gas at density $r_s=2$.
The present model is compared with the Quantum Monte Carlo (QMC)
data from Ref.~\protect\onlinecite{OHB} and with the
result obtained from the local-field factors of
Richardson and Ashcroft\protect\cite{RA} (RA).}
\label{fig_spinres}
\end{figure}

At metallic densities, we used the spin-resolved
$\ec$ for the $\zeta=0$ gas from Ref.~\onlinecite{GSB},
and the RPA spin resolution for the long-range part.
In Fig.~\ref{fig_spinres}, we report our results for
$g_c^{\ud}$ and $g_c^{\uu}=2g_c-g_c^{\ud}$
at $r_s=2$, together with the QMC data of Ref.~\onlinecite{OHB},
and with the values that we have obtained from
the Richardson and Ashcroft (RA) local-field factors.\cite{RA}
We see that our result is reasonable, but does not accurately agree
with the QMC data. In this respect, the RA results are much better
for $u/r_s \gtrsim 0.7$, while they blow up in the short-range
part, since they do not satisfy the Pauli principle in real space.
As said, the spin resolution is very delicate, so that
an analytic model is very difficult to build up. The best
analytic representation of $g_c^{\ud}$ and $g_c^{\uu}$ at metallic
densities is probably the one of Ref.~\onlinecite{GSB}, which
was built to interpolate the QMC data of Ref.~\onlinecite{OHB}
accurately.

\section{Wavevector analysis of the kinetic energy of correlation}
\label{sec_tc}
\begin{figure}
\includegraphics[width=7cm]{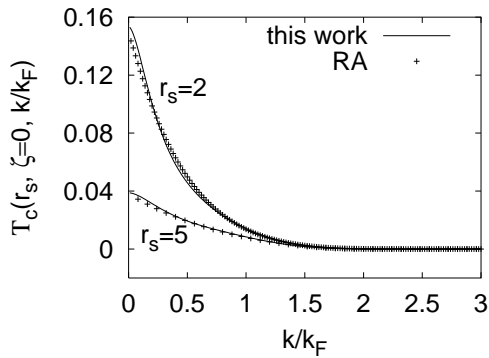} 
\caption{Wavevector analysis of the kinetic energy of correlation
at $r_s=2$ and $r_s=5$ for the paramagnetic gas. The function
${\mathcal T}_c(r_s,\zeta,k)$ is defined in Eq.~(\protect\ref{eq_defT}).
The present work is compared with the result obtained from the 
Richardson and Ashcroft\protect\cite{RA} (RA) local-field factor.} 
\label{fig_tc}
\end{figure}

Wavevector analysis\cite{LP} is usually a study of the static
structure factor of Eq.~(\ref{eq_Sxcdef}).
The wavevector analyses of the correlation energy $\epsilon_c$
and of the potential energy of correlation $v_c$ have often been 
reported,\cite{GSB,LGP,TGPM} while the kinetic energy of correlation 
$t_c$ is
much less studied. We can decompose $t_c(r_s,\zeta)$ into 
contributions from different wavevectors of a density fluctuation,
\be
t_c(r_s,\zeta)=\tfrac{3}{2}\int_0^{\infty} dq\,q^2\,\mathcal{T}_c(r_s,\zeta,q),
\label{eq_tcq}
\end{equation}
where $q=k/k_F$. Since $t_c = \epsilon_c - v_c$, the
wavevector analysis of $t_c$ is just the difference between those for
$\epsilon_c$ and $v_c$:
\be
\mathcal{T}_c(r_s,\zeta,q)=\frac{2k_F}{3\pi}\frac{[\overline{S}_c(r_s,\zeta,q)-
S_c(r_s,\zeta,q)]}{q^2}.
\label{eq_defT}
\end{equation}
The small-$q$ limit of $\mathcal{T}_c$ can be obtained by the plasmon
sum rule, 
\be
\mathcal{T}_c(r_s,\zeta,q\to 0)=\tfrac{\sqrt{3}}{4}\,r_s^{-3/2}+O(q^2),
\label{eq_Tsmallq}
\end{equation}
and its leading term is independent of $\zeta$, as expected from
Eq.~(\ref{eq_plas}).
To write down the large-$q$ limit of $\mathcal{T}_c$ we need to expand
$\overline{S}_c$ and $S_c$ for large arguments. We know 
that\cite{Kimball2,Kimball,GSB,GSB2}
\be
S_c(r_s,\zeta,q \to \infty)=-\frac{4}{3\pi k_F}\frac{2 g_{xc}(r_s,\zeta,u=0)}
{q^4}+O(q^{-6}),
\end{equation}
from which we can also obtain the large-$q$ limit of $\overline{S}_c$,
\be
\overline{S}_c(r_s,\zeta,q \to \infty)=\frac{\overline{\gamma}}{q^4}
+O(q^{-6}),
\end{equation}
where
\be
\overline{\gamma}=-\frac{8}{3\pi}\left(\frac{4}{9\pi}\right)^{1/3}
\frac{1}{r_s}\int_0^{r_s} r_s'\,g_{xc}(r_s',\zeta,u=0)\,dr_s'.
\end{equation}
Through the cusp condition of Eq.~(\ref{eq_cusp}), we see that the
large-$q$ limit of $\overline{S}_c$ is determined by the coefficient
of $u/r_s$ in the small-$u$ expansion of $\overline{g}_c$,
$\overline{a}_1(r_s,\zeta)$
[see Eqs.~(35) and~(37) of Ref.~\onlinecite{GP}], related to
$c_2(r_s,\zeta)$ of Eq.~(\ref{eq_c2})
by $\overline{a}_1=(9\pi/4)^{1/3}c_2/\phi$.
We thus have
\be
\mathcal{T}_c(r_s,\zeta,q\to \infty)=\frac{8}{9\pi^2}
\frac{\left[2g_{xc}(r_s,\zeta,u=0)-2\frac{\overline{a}_1(r_s,\zeta)}{r_s}
\right]}{q^6}.
\label{eq_Tlargeq}
\end{equation}
In Fig.~\ref{fig_tc}, we report $\mathcal{T}_c$ for the $\zeta=0$ gas,
for two different densities, $r_s=2$ and $r_s=5$. We clearly see that 
the small wavector contribution to $t_c$ comes from the kinetic energy
of the long-wavelength zero-point plasmons, and that the decay of 
the plasmon contribution with increasing wavevector $k$ is gradual. 
The corresponding result from the Richardson and Ashcroft 
local field factor\cite{RA} is also shown.
The Richardon-Ashcroft model gives a good description
of plasmon dispersion and damping.\cite{TSS}

It is also interesting to compare $\mathcal{T}_c$ with the 
decomposition of the kinetic energy of correlation into
contributions from different wavevectors of a quasi-electron. For 
$\zeta=0$, we can write
\be
t_c(r_s,\zeta=0)=\tfrac{3}{2}\int_0^{\infty}dq\,q^2\,n_c(r_s,\zeta=0,q)\,
(k_F\,q)^2.
\end{equation}
Here $n_c$ is the correlation contribution to the momentum distribution,
$n_c(q)=n(q)-n_0(q)$, and $n_0$ is the Fermi step function.
The leading term in the small-$q$ expansion of $k_F^2 q^2 n_c$ is 
proportional to $q^2$, and is thus rather different from 
the corresponding behavior of $\mathcal{T}_c$, Eq.~(\ref{eq_Tsmallq}).
On the other hand, in the large-$q$ limit we have\cite{Kimball}
\be
k_F^2q^2n_c(r_s,\zeta=0,q\to \infty)=
\frac{8}{9\pi^2}\frac{g_{xc}(r_s,\zeta=0,u=0)}{q^6},
\label{eq_nclargeq}
\end{equation}
a behavior very similar to Eq.~(\ref{eq_Tlargeq}). 
This is not surprising, since the large-$q$ limits of
both $S_c$ and $n_c$ are determined by the downward-pointing kinks in the
many body wavefunction [producing the cusp of Eq.~(\ref{eq_cusp})] 
which occur whenever 
two electrons of antiparallel spin come together.
In the $r_s\to 0$ limit, Eqs.~(\ref{eq_Tlargeq}) and~(\ref{eq_nclargeq}) 
become equal. A study of the equations linking $S_c$ and
$n_c$ from the point of view of density matrix functional
theory is reported in Ref.~\onlinecite{GZ}.

\section{Conclusions and future directions}
\label{sec_conc}
The known exact constraints summarized in Sec.~\ref{sec_prop}, plus the
random phase approximation for long-range ($u\to \infty$) correlation, the
extended Overhauser model\cite{GP} for short-range ($u\to 0$) correlation, 
and the correlation energy $\ec(r_s,\zeta)$, suffice to determine the pair
distribution function $g_{xc}(r_s,\zeta,k_Fu)$ of a uniform electron gas over
the whole density range, including the high-density ($r_s\to 0$) and
low-density ($r_s\to\infty$) limits, apart from energetically-unimportant
long-range oscillations.  The analytic formulas we have so constructed
for the coupling-constant-averaged
$\gxcav$ should be useful for further developments and applications of
density-functional approximations for the exchange-correlation energy of a
non-uniform density.\par

For metallic densities ($2<r_s<10$) with $\zeta=0$ or 1, our $g_{xc}$ is in
good agreement with Quantum Monte Carlo.\cite{cepald,OHB}  In the same density
range for $\zeta=0$, it also agrees with the $g_{xc}$ we have calculated 
from the Richardson-Ashcroft\cite{RA} dynamic local-field factor, 
except near $u=0$ where the Richardson-Ashcroft model 
was found to break down (although this model seems to describe the long-range
oscillations correctly).  In the $r_s\to 0$ limit for small
$k_Fu$, our $g_{xc}$ agrees with the results of perturbation theory to zero-th
(exchange) or first order\cite{Rassolov,reply} in the electron-electron 
interaction.
The static structure factor $S_{xc}$ is also modelled accurately,
neglecting the non-analytic structure of the exact $S_{xc}$ at 
$k=2k_F$ arising from long-range oscillations.\cite{GZ}\par

Our formulas can also be used to spin-resolve $g_{xc}$ into $\uu$,
$\dd$, and $\ud$ components (Sec.~\ref{sec_spinres}), when the spin 
resolution of $\ec$ is known (as it is in the high- and low-density 
limits). The additional information in the spin resolution
might well be used to construct more accurate density functionals for the
correlation energy.\par

      We have also examined two physically-different wavevector analyses
of the kinetic energy of correlation in the uniform electron gas, finding
them the same only in the limits of large wavevector and high density.
We have also found that the decay of the plasmon contribution with 
increasing wavevector $k$ is gradual.\par

In the future, it may be possible to construct the correlation
energy $\ec(r_s,\zeta)$ and its spin resolution $\epsilon_c^{\sigma \sigma'}
(r_s,\zeta)$ directly by interpolation between known limits,\cite{mike} 
without using {\em any} Monte Carlo or other data.  
This development would probably not give us a better $\epsilon_c(r_s,\zeta)$
than we already have, but would provide the first spin resolution over the
whole range of $r_s$ and $\zeta$; it would also show that the known exact
constraints are by themselves sufficient to determine $g_{xc}$.
The extended Overhauser
model\cite{GP} might be evaluated for $\zeta$ different from zero, to test and
refine the spin-scaling relations used in Ref.~\onlinecite{GP}.  The extended
Overhauser model can also be made more selfconsistent.\cite{DPAT}
\par
A small {\tt FORTRAN77} subroutine which numerically evaluates 
our $\gcav$ [Eq.~(\ref{eq_base})] can be downloaded
at {\tt http://axtnt2.phys.uniroma1.it/PGG/elegas.html}.

\section*{Acknowledgments}
This work was supported by the Fondazione Angelo
Della Riccia (Firenze, Italy), the MURST (the Italian 
Ministry for University, Research and Technology) through COFIN99,
and by the U.S. National Science Foundation 
under grants DMR-9810620 and DMR-0135678. 
\appendix
\section{Nonoscillatory exchange hole in reciprocal space}
\label{app_exck}
In reciprocal space, the exchange-only static structure factor
is equal to
\begin{eqnarray}
S_x(\zeta,k/k_F)& = & 1+\frac{2}{3\pi}\Biggl[(1+\zeta)\tilde{J}
\left(\frac{k}{k_F(1+\zeta)^{1/3}}\right)+ \nonumber \\
& & (1-\zeta)\tilde{J}
\left(\frac{k}{k_F(1-\zeta)^{1/3}}\right)\Biggr],
\end{eqnarray}
where $\tilde{J}(k)$ is defined by Eq.~(\ref{eq_Jk}). From
our parametrization of $\langle J(y)\rangle$ [Eq.~(\ref{eq_J})]
we obtain
\bea
\tilde{J}(k) & = & \frac{9\pi}{16} k \left[1-{\rm erf}\left(\frac{k}{2\sqrt{A_x}}
\right)\right]-\frac{3\sqrt{\pi}}{32}e^{-\frac{k^2}{4A_x}}\times
\nonumber \\
& & \left(9\sqrt{A_x}+\frac{k^2-6A_x}{4\sqrt{A_x}}\right)+\frac{\sqrt{\pi}}{4}
e^{-\frac{k^2}{4D_x}}\Biggl[\frac{B_x}{D_x^{3/2}} + \nonumber \\
& & \frac{C_x(6D_x-k^2)}{4D_x^{7/2}}+\frac{E_x(60D_x^2-20D_xk^2+k^4)}{16D_x^{11/2}}+
\nonumber \\
& & \frac{F_x(840 D_x^3-420 D_x^2k^2+42 D_xk^4-k^6)}{64 D_x^{15/2}}\Biggr].
\eea

\section{Long-range correlation hole in reciprocal space}
\label{app_fz}
The function $f(z,0)$ corresponding to our Eq.~(\ref{eq_newf1}) is
\bea
f(z,0)& = & \frac{1}{2zb^8}\Big\{a_0- \frac{e^{-bz}}{48}\big[48 a_0
+(33a_0b-3a_1b^3 \nonumber \\
& & -3a_2b^5-15a_3b^7)z+(9a_0b^2-3a_1b^4-\nonumber \\
& & 3a_2b^6+9a_3b^8)z^2+
 (a_0b^3-a_1b^5+a_2b^7 \nonumber \\
& & -a_3b^9)z^3\big]\Big\}-\frac{b_2}{6\pi\,z}\frac{\partial^3{\cal I}(b,z)}
{\partial (b^2)^3},
\label{eq_SFTf1}
\eea 
where
\be
 {\cal I}(b,z)   =  \frac{1}{2b}\left[e^{bz}E_1(bz)
-e^{-bz}E_1(-bz)\right],
\end{equation}
and, with $x>0$,
\begin{eqnarray}
& & E_1(x)=\int_x^{\infty}\frac{e^{-t}}{t}dt
\nonumber \\
& & E_1(-x)=-{\rm Ei}(x)=-{\rm PV}\left(\int_{-\infty}^x\frac{e^t}{t}dt\right).
\nonumber 
\end{eqnarray}
Here PV means the Cauchy principal value integral.\cite{abra}\par

The parameter values which satisfy Eqs.~(\ref{eq_fzsmallz}) and
(\ref{eq_fzlargez}),
give rise to a zero coefficient for the $z^3$ term in the small-$z$
expansion of $f(z,0)$, and accurately fit our RPA data,\cite{WP} are
\bea
a_0 & = & 2b^8C_0 \nonumber \\
a_1 & = & \frac{6b^3}{\pi^2}\big[\pi^2 C_0 b^3+78 b-256\sqrt{3}\big]
\nonumber \\
a_2 & = & 48 b^2/\pi^2 \nonumber \\
a_3 & = & 12/\pi^2 \nonumber \\
b_2 & = & \frac{3b^4}{\pi}\big[96\sqrt{3}-36b-b^3C_0\pi^2\big] \nonumber \\
C_0 & = & -2(1-\ln 2)/\pi^2 \nonumber \\
b & = & 7.8. \nonumber
\eea

\section{Analytic expressions for the functions $S(\alpha)$, $P(\alpha)$ and
$R(\alpha)$}
\label{app_SPR}
The three functions of Eqs.~(\ref{eq_Sa}), (\ref{eq_Pa}) and~(\ref{eq_Ra}), 
which enter our model for $\gcav$, are given by a linear combination of
integrals of the kind
\bea
{\cal I}_m^n(\alpha)&  = & \int_0^{\infty}\frac{x^ne^{-x^2}}{[(\alpha x)^2+b^2]^m}
dx \label{eq_Imn}\\
{\cal I}_m^{-1}(\alpha)&  = &\int_0^{\infty}\frac{1-e^{-x^2}}{x[(\alpha x)^2+b^2]^m}
dx.\label{eq_I-1n}
\eea
We obtain for $S(\alpha)$, $P(\alpha)$ and $R(\alpha)$:
\bea
S(\alpha) & = & a_0 {\cal I}_4^0(\alpha)+(a_0+a_1\alpha^2){\cal I}_4^2(\alpha)+
(\tfrac{1}{2}a_0+a_1\alpha^2+ \nonumber \\
& & a_2\alpha^4){\cal I}_4^4(\alpha)+(\tfrac{1}{2}a_1\alpha^2+a_2\alpha^4
+a_3\alpha^6){\cal I}_4^6(\alpha)+ \nonumber \\
& & (\tfrac{1}{2}a_2\alpha^4+a_3\alpha^6){\cal I}_4^8(\alpha)+\tfrac{1}{2}
a_3\alpha^6{\cal I}_4^{10}(\alpha)+b_2\alpha\times
\nonumber \\
& &  \left[{\cal I}_4^1(\alpha)+{\cal I}_4^3(\alpha)+
\tfrac{1}{2}{\cal I}_4^5(\alpha)\right] \\
P(\alpha) & = & a_0 {\cal I}_4^2(\alpha)+(a_0+a_1\alpha^2){\cal I}_4^4(\alpha)+
(\tfrac{1}{2}a_0+a_1\alpha^2+ \nonumber \\
& & a_2\alpha^4){\cal I}_4^6(\alpha)+(\tfrac{1}{2}a_1\alpha^2+a_2\alpha^4
+a_3\alpha^6){\cal I}_4^8(\alpha)+ \nonumber \\
& & (\tfrac{1}{2}a_2\alpha^4+a_3\alpha^6){\cal I}_4^{10}(\alpha)+\tfrac{1}{2}
a_3\alpha^6{\cal I}_4^{12}(\alpha)+b_2\alpha\times
\nonumber \\
& &  \left[{\cal I}_4^3(\alpha)+{\cal I}_4^5(\alpha)+
\tfrac{1}{2}{\cal I}_4^7(\alpha)\right] \\
R(\alpha) & = & a_0 {\cal I}_4^{-1}(\alpha)-
(a_0+a_1\alpha^2){\cal I}_4^1(\alpha)-
(\tfrac{1}{2}a_0+a_1\alpha^2+ \nonumber \\
& & a_2\alpha^4){\cal I}_4^3(\alpha)-(\tfrac{1}{2}a_1\alpha^2+a_2\alpha^4
+a_3\alpha^6){\cal I}_4^5(\alpha)- \nonumber \\
& & (\tfrac{1}{2}a_2\alpha^4+a_3\alpha^6){\cal I}_4^7(\alpha)-\tfrac{1}{2}
a_3\alpha^6{\cal I}_4^9(\alpha)-b_2\alpha\times
\nonumber \\
& & \left[{\cal I}_4^0(\alpha)+
{\cal I}_4^2(\alpha)+
\tfrac{1}{2}{\cal I}_4^4(\alpha)\right]+\nonumber \\
& & \frac{2a_1+a_2b^2+2a_3b^4}{12b^6}+b_2\frac{5}{32}\frac{\pi}{b^7},
\eea
where $a_0$, $a_1$, $a_2$, $a_3$, $b_2$ and $b$ are given in 
Appendix~\ref{app_fz}.
The integrals of the kind~(\ref{eq_Imn}) can be written as
\bea
{\cal I}_m^n(\alpha)& = & \tilde{{\cal I}}_m^n(\alpha/b)/b^{2n}, \\
\tilde{{\cal I}}_m^n(r)& = & \int_0^{\infty}\frac{x^ne^{-x^2}}{[(rx)^2+1]^m}dx,
\eea
and starting from
\bea
\tilde{{\cal I}}_1^0(r) & = & \frac{\pi}{2r^2}e^{1/r^2}\left[1-{\rm erf}\left(
\frac{1}{r}\right)\right] \\
\tilde{{\cal I}}_1^1(r) & = & \frac{1}{2r^2}e^{1/r^2}E_1\left(\frac{1}{r^2}
\right),
\eea
can be computed by differentiation with respect to $\alpha$ and $b$.
The integrals of the kind~(\ref{eq_I-1n}) can be also obtained
by differentiation with respect to $b$ of
\be
{\cal I}_1^{-1}(\alpha)=\frac{e^{b^2/\alpha^2}E_1(b^2/\alpha^2)
+\ln(b^2/\alpha^2)+\gamma}{2b^2},
\end{equation}
where $\gamma=0.5772156649$.


\begin{thebibliography}{100}
\bibitem{nota} In Refs.~\onlinecite{GSB,GSB2,comment} the RPA 
spin-resolution is assumed exact for the long-range part 
of $g_{xc}$ at all densities. Further studies by one of us (P.G.-G.)
showed that this is true only when $r_s \to 0$, but
with small deviations as $r_s$ increases, so that the assumption made
is still quite accurate at $r_s\sim 10$.
\bibitem{nota2} This assumption is reasonable, but could be 
inexact. In the low-density limit, the electron gas becomes 
$\zeta$-independent, since the statistics is irrelevant with respect
to Coulomb repulsion, but magnetic order can be present, so that
$g_{xc}^{\ud}$ may differ from $g_{xc}^{\uu}$ for $u/r_s\gtrsim 1$.
A close comparison of our Figs.~\ref{fig_LD} and~\ref{fig_gudLD} 
for $r_s = 10^5$  and $\zeta = 0$ seems to show this effect.
(For $u/r_s \lesssim 1$ both  $g_{xc}^{\ud}$ and $g_{xc}^{\uu}$ tend to
zero in the strongly correlated limit.)
\end{thebibliography}
\end{document}